\documentclass[a4paper,12pt]{article}
\usepackage{feynmp-auto,expdlist}
\usepackage{amsmath, amsfonts}
\usepackage{graphicx,arydshln}
\usepackage{enumerate}
\usepackage{hyperref}
\usepackage{latexsym}
\usepackage{dsfont}
\usepackage{hepnicenames}
\usepackage{enumerate}
\usepackage{soul}
\usepackage[normalem]{ulem}
\usepackage{comment}

\newcommand{\footnoten}[1]{}

\usepackage[font={small},textfont={it}]{caption} % make captions small

\def\tit{{\tilde{t}}}
\def\rt{{\tilde{r}}}
\def\tit{{{t}}}
\def\rt{{{r}}}

\DeclareMathOperator{\atanh}{atanh}

\usepackage{units}
\usepackage{mathrsfs,graphicx,rotating,amsmath,amsfonts,mathtools,booktabs,amssymb,wasysym}
\usepackage{hyperref}\usepackage{slashed}
\usepackage[nosort]{cite}
\usepackage[table,xcdraw,dvipsnames]{xcolor}
\usepackage{bm}
\usepackage{multirow,multicol}
\hypersetup{colorlinks,bookmarksopen,bookmarksnumbered,
	linkcolor=blus,pdfstartview=FitH,urlcolor=rossos,citecolor=verde}
\allowdisplaybreaks

\newcommand{\HI}{H_{\rm infl}}
\newcommand{\myfootnote}[1]{}
\newcommand{\myomit}[1]{{\color{gray}#1}}
\renewcommand{\myomit}[1]{}

\renewcommand{\[}{\left[}

\newcommand{\bP}{\bar{M}_{\rm Pl}}

\def\Lag{\mathscr{L}}

\newcommand{\mio}[1]{}

\def\bpm{\begin{pmatrix}}
	\def\epm{\end{pmatrix}}

\usepackage{mathrsfs}

\newcommand{\fig}[1]{~\ref{fig:#1}}
\newcommand{\sfrac}[2]{#1/#2}

\allowdisplaybreaks
\usepackage{multicol}
\usepackage{color}
\definecolor{rosso}{cmyk}{0,1,1,0.4}
\definecolor{rossos}{cmyk}{0,1,1,0.55}
\definecolor{rossoc}{cmyk}{0,1,1,0.2}
\definecolor{blu}{cmyk}{1,1,0,0.3}
\definecolor{blus}{cmyk}{1,1,0,0.6}
\definecolor{bluc}{cmyk}{1,1,0,0.1}
\definecolor{verde}{cmyk}{0.92,0,0.59,0.25}
\definecolor{verdec}{cmyk}{0.92,0,0.59,0.15}
\definecolor{verdes}{cmyk}{0.92,0,0.59,0.4}

\newcommand{\bp}{\bar{M}_{\rm Pl}}
\newcommand{\eq}[1]{~{\rm (\ref{eq:#1})}}

\newcommand{\GeV}{\,{\rm GeV}}

\def\circa#1{\,\raise.3ex\hbox{$#1$\kern-.75em\lower1ex\hbox{$\sim$}}\,}

\newcommand{\beq}{\begin{equation}}
\newcommand{\eeq}{\end{equation}}

\newcommand{\bea}{\begin{eqnarray}}
\newcommand{\eea}{\end{eqnarray}}
\newcommand{\be}{\begin{equation}}
\newcommand{\ee}{\end{equation}}
\font\tenrsfs=rsfs10 at 12pt
\font\sevenrsfs=rsfs7
\font\fiversfs=rsfs5
\newfam\rsfsfam
\textfont\rsfsfam=\tenrsfs
\scriptfont\rsfsfam=\sevenrsfs
\scriptscriptfont\rsfsfam=\fiversfs

\newsavebox\MBox

\renewenvironment{thebibliography}[1]
{\begin{multicols}{2}[\section*{\refname}]%
		\@mkboth{\MakeUppercase\refname}{\MakeUppercase\refname}%
		\list{\@biblabel{\@arabic\c@enumiv}}%
		{\settowidth\labelwidth{\@biblabel{#1}}%
			\leftmargin\labelwidth
			\advance\leftmargin\labelsep
			\@openbib@code
			\usecounter{enumiv}%
			\let\p@enumiv\@empty
			\renewcommand\theenumiv{\@arabic\c@enumiv}}%
		\sloppy
		\clubpenalty4000
		\@clubpenalty \clubpenalty
		\widowpenalty4000%
		\sfcode`\.\@m}
	{\def\@noitemerr
		{\@latex@warning{Empty `thebibliography' environment}}%
		\endlist\end{multicols}}

\renewcommand{\L}\Lag

\def\circa#1{\,\raise.3ex\hbox{$#1$\kern-.75em\lower1ex\hbox{$\sim$}}\,}
\makeatletter

\font\ital=cmu10

\def\hhref#1{\href{http://arxiv.org/abs/#1}{arXiv:#1}}
\usepackage{xstring}
\newcommand{\hhrefq}[1]{\IfSubStr{#1}{:}{\href{http://inspirehep.net/search?ln=en&ln=en&p=#1&of=hb&action_search=Search&sf=&so=d&rm=&rg=25&sc=0}{InSpire:#1}}{\hhref{#1}}}

\def\art{\@ifnextchar[{\eart}{\oart}}
\def\eart[#1]#2#3#4#5#6{{\rm #2}, {\em #3 \bf #4} {\rm (#6) #5} ({\em #1})}
\def\article{\@ifnextchar[{\earticle}{\oarticle}}
\def\oarticle#1#2#3#4#5#6{{\rm #1}, {\ital `#6'}, {\rm #2 #3 (#5) #4}}
\def\earticle[#1]#2#3#4#5#6#7{{\rm #2}, {\ital `#7'}, {\rm #3 #4 (#6) #5}  [\hhrefq{#1}]}
\def\hepart[#1]#2{{\rm #2, \sl#1}}
\def\heparticle[#1]#2#3{#2, {\ital `#3'} [\hhrefq{#1}]}
\newcommand{\doi}[1]{\href{http://dx.doi.org/#1}{[link]}}

\newcommand{\hhrefqq}[1]{\IfBeginWith{#1}{10.}{\href{https://doi.org/#1}{doi:#1}}{\hhrefq{#1}}}
\def\earticle[#1]#2#3#4#5#6#7{{\rm #2}, {\ital `#7'}, {\rm #3 #4 (#6) #5}  [\hhrefqq{#1}]}
\renewenvironment{thebibliography}[1]
{\begin{multicols}{2}[\section*{\refname}]%
		\@mkboth{\MakeUppercase\refname}{\MakeUppercase\refname}%
		\list{\@biblabel{\@arabic\c@enumiv}}%
		{\settowidth\labelwidth{\@biblabel{#1}}%
			\leftmargin\labelwidth
			\advance\leftmargin\labelsep
			\@openbib@code
			\usecounter{enumiv}%
			\let\p@enumiv\@empty
			\renewcommand\theenumiv{\@arabic\c@enumiv}}%
		\sloppy
		\clubpenalty4000
		\@clubpenalty \clubpenalty
		\widowpenalty4000%
		\sfcode`\.\@m}
	{\def\@noitemerr
		{\@latex@warning{Empty `thebibliography' environment}}%
		\endlist\end{multicols}}

\newcounter{alphaequation}[equation]
\def\thealphaequation{\theequation\hbox to
	0.6em{\hfil\alph{alphaequation}\hfil}}
\def\eqnsystem#1{
	\def\@eqnnum{{\rm (\thealphaequation)}}
	\def\@@eqncr{\let\@tempa\relax \ifcase\@eqcnt \def\@tempa{& & &} \or
		\def\@tempa{& &}\or \def\@tempa{&}\fi\@tempa
		\if@eqnsw\@eqnnum\refstepcounter{alphaequation}\fi
		\global\@eqnswtrue\global\@eqcnt=0\cr}
	\refstepcounter{equation} \let\@currentlabel\theequation \def\@tempb{#1}
	\ifx\@tempb\empty\else\label{#1}\fi
	\refstepcounter{alphaequation}
	\let\@currentlabel\thealphaequation
	\global\@eqnswtrue\global\@eqcnt=0 \tabskip\@centering\let\\=\@eqncr
	$$\halign to \displaywidth\bgroup \@eqnsel\hskip\@centering
	$\displaystyle\tabskip\z@{##}$&\global\@eqcnt\@ne
	\hskip2\arraycolsep\hfil${##}$\hfil& \global\@eqcnt\tw@\hskip2\arraycolsep
	$\displaystyle\tabskip\z@{##}$\hfil
	\tabskip\@centering&\llap{##}\tabskip\z@\cr}
\def\endeqnsystem{\@@eqncr\egroup$$\global\@ignoretrue} \makeatother

\definecolor{Gray}{gray}{0.95}

\def\bal#1\eal{\begin{align}#1\end{align}}

\begin{document}
\begin{center}  
{\huge\bf\color{rossos} Higgstory repeats itself} \\[3ex]
{\bf\large Alessandro Strumia$^{a}$ and Nikolaos Tetradis$^{b,c}$}\\[2ex]
{\it $^a$ Dipartimento di Fisica, Universit\`a di Pisa, Italia}\\
{\it $^b$ Department of Physics, National and Kapodistrian University of Athens, Greece}\\
{\it $^c$ } {\em CERN, Theoretical Physics Department, Geneva, Switzerland}\\[3ex]

{\large\bf Abstract}\begin{quote}
We consider a scalar potential with two minima, one of which is arbitrarily deep,  
such as could be the case for the Higgs potential in the Standard Model.
A recent calculation within the thin-wall approximation~\cite{Riotto} 
concludes that regions in which the scalar field takes values beyond the top of 
the potential barrier are forced by gravity to collapse, while they
remain hidden behind a black hole horizon.
We show that the thin-wall approximation is not applicable to this problem.
We clarify the issue through numerical and analytical solutions to the field equations 
of the gravity-scalar system.
We find that regions around the deeper minimum expand, and would thereby 
engulf the Universe in post-inflationary cosmology.
We also show that black holes with Higgs hair are unstable.
Even though the physics of the true vacuum is different, our final conclusion replicates 
the earlier `Higgstory' paper~\cite{Higgstory}. 
\end{quote}
\end{center}
\tableofcontents

\section{Introduction}
A decade ago the Higgs boson mass was measured~\cite{1207.7235,1207.7214},
and its value $M_h \approx 125.1\GeV$
implied the possible instability of the Standard Model Higgs potential  at large field values $h > h_{\rm top} \sim 10^{10}\GeV$~\cite{1112.3022,1307.3536,1507.08833}.
This instability would have important cosmological implications~\cite{Higgstory,1112.3022,1210.6987,1301.2846,1503.05193,1605.04974,1606.00849,1608.02555,1608.08848,1706.00792,1809.06923,2011.03763}.
Establishing if the SM Higgs potential is really unstable needs a more accurate determination of the top quark mass,
a task that seems feasible only at a future lepton collider at the $t\bar t$ threshold~\cite{FCC:2018evy,CEPC}, possibly in the LEP tunnel~\cite{2203.17197}.

\smallskip

Motivated by this possible instability, we reconsider a more general cosmological issue:
if a scalar $h$ sits at the local minimum $h=h_{\rm false}$ of a potential $V(h)$
that also has a deeper minimum at $h=h_{\rm min}$
(so that $V_{\rm min} \equiv V(h_{\rm min})< V(h_{\rm false}) \equiv V_{\rm false}  $)
beyond a potential barrier at $h=h_{\rm top}$, 
what is the fate of space-time regions where the field $h$ acquires values $h > h_{\rm top}$?
%such that $V(h) $?
As the field tends to roll down towards the deeper minimum, 
one expects that such regions would grow in space, assuming that they are large enough
for the potential energy to dominate over gradient energy.

The situation is less clear when gravity is taken into account, as regions 
with negative energy density tend to undergo an AdS-like gravitational collapse
towards an uncertain final state. 
This problem was studied in~\cite{Higgstory} (`Higgstory' paper), 
where such regions were approximated as thin-wall spherical bubbles with $h=h_{\rm min}$ 
inside and $h=h_{\rm false}$ outside.
It was found that  an observer inside the bubble experiences an AdS crunch, 
while an outside observer sees the bubble expand.
When this picture is applied to the SM case, it implies that 
expanding bubbles would engulf the whole 
universe once inflation ends.
This is not observed, implying bounds on cosmology, in particular on the 
inflationary Hubble scale $\HI $~\cite{Higgstory, 
1606.00849,1608.02555,1706.00792,1809.06923,2011.03763}.

A recent paper~\cite{Riotto} reaches a different conclusion 
by extending the thin-wall approximation used in~\cite{Higgstory}
to analyze also the earlier phase during which $h$ falls down the potential towards 
$h_{\rm min}$. This phase was not studied in \cite{Higgstory}. However, for a 
very deep true vacuum, or an unbounded potential, it may play a crucial role 
in the evolution of the system.
The authors of~\cite{Riotto} correctly point out that the energy 
density $\rho \approx K + V$  inside the bubble  decreases towards zero
(as $h$ rolls down the potential, 
while the kinetic energy $K\approx \dot h^2/2$ is red-shifted away by the Hubble friction),
until the local scale factor stops expanding and starts to contract.
Subsequently, $K$ gets blue-shifted and becomes dominant
during the rapid fall of $h$ towards the true minimum, with
a singularity developing if the minimum is very deep.
However, \cite{Riotto} also claims that, as a result, 
the thin wall of the bubble starts moving inward until it disappears, hidden beyond
a static black hole horizon. The exterior space-time remains largely unaffected, 
apart from the appearance of an isolated black hole.

This is puzzling, as physical systems with stochastic fluctuations 
are commonly expected to evolve towards vacua with lower energy
and eventually find their way to the true vacuum. 
If this scenario were true, it would have a big impact on more general similar situations, 
such as spontaneous vacuum decay to AdS vacua, multiverse inflationary dynamics,
ultra-high-energy collisions of cosmic rays~\cite{Hut:1983xa,Rubakov:1991fb,hep-ph/9703256,hep-ph/9704431,hep-ph/9910333}, 
or even at colliders and possibly more general Higgs engineering.
We thereby re-examine this issue. Even though we confirm the analysis of \cite{Riotto}
for the initial stage of the evolution, 
our final results replicate the original Higgstory \cite{Higgstory}
conclusion that the bubbles can expand.

The crucial point is that the simplifying thin-wall assumption 
---  an AdS bubble with constant $h$ at the minimum  $h_{\rm min}$ inside ---
was generalized in~\cite{Riotto} to an assumption for constant field $h\neq h_{\rm min}$ 
over the whole interior.
However, assuming a constant field value
away from the potential minimum is an unphysical `rigidity' assumption that 
artificially links the inner crunch to the wall boundary,
forcing the contraction and disappearance of the whole bubble.
In section~\ref{bubble} we show that the thin-wall approximation is not applicable to 
the general situation with $h\neq h_{\rm min}$.
What happens instead is that even an initially thin wall gets stretched on both sides
(in a sense falling down towards either minimum), thus
becoming thick.
The configuration looks like a `sinkhole'. %also dubbed `portal to hell'.

In order to establish the physical mechanism at work,
we derive in section~\ref{sinkhole} the full equations of the gravity-scalar 
system,
and solve them in multiple ways: we find one analytic solution in one special case,
and numerical solutions in generic cases, using two different forms of
the dynamical metric (FRW-like or static-like).
All solutions lead to the same general conclusion:
the process is similar to a gravitational collapse, 
where an inner region contracts  without affecting substantially the
outer region, which keeps expanding
because the two get causally disconnected and/or because information about 
the collapse does not propagate fast enough.
A black hole forms, but it sits in the true vacuum and thereby is not static, as 
it accretes energy.
In section~\ref{hair} we explore related configurations that can be described as
static black holes 
with Higgs hair, finding that they are unstable.

Conclusions are given in section~\ref{concl}.

%\begin{figure}[t]
%$$\includegraphics[width=0.45\textwidth]{figs/finfoutsampledSMneg}\qquad
%\includegraphics[width=0.45\textwidth]{figs/finfoutsampledSMpos}$$
%\caption{\it Two samples of $f(r)$ in thin wall approximation. 
%The first is an AdS bubble inside flat space.
%The second is a BH inside an AdS bubble inside dS.
%Both have $\sigma$ so large that $M_{\rm out}>0$ but this makes the discontinuity uglier.
%AdS bubbles with large radius naturally have $M_{\rm out}<0$ dominated by the large inside volume,
%and not compensated by the small tension. In such a case no critical bubble exists: bubbles either expand or contract.
%\label{fig:finfoutsample}}
%\end{figure}

%\begin{figure}[t]
%$$$$
%\caption{\it }
%\end{figure}

\section{The form of the bubbles}\label{bubble}

We consider a scalar field $h$ coupled to gravity.
Without loss of generality, a non-minimal scalar coupling to gravity can be removed by a field redefinition, and 
the system can be studied in the Einstein frame, where both 
scalar and gravity have canonical kinetic terms.
The action then is
\beq S = \int d^4x \sqrt{|\det g|} 
\left[\frac{\bp^2}{2} R - \frac12 g^{\mu\nu}(\partial_\mu h)(\partial_\nu h) - V(h)\right],\eeq
where we use the convention $(-,+,+,+)$ for the signature of the metric.
We assume that the scalar potential $V(h)$ has a false minimum with $V_{\rm false} = V(h_{\rm false})$ and a
true minimum $V_{\rm min}=V(h_{\rm min})$, possibly very deep.
They are separated by a potential barrier $V_{\rm top}=V(h_{\rm top})$.
For example, we can consider potentials with $h_{\rm false}=0$ of the form
\beq \label{eq:V}V = \frac{3\HI ^2}{8\pi G} + m^2 \frac{h^2}{2} + \lambda \frac{h^4}{4} + \frac{h^6}{\Lambda^2}\eeq
with $\lambda<0$ and $G = 1/M_{\rm Pl}^2 =1/8\pi\bp^2$.
 The inflationary energy can be provided by some other 
scalar, whose nature is not relevant here.
We want to establish if a region where $h > h_{\rm top}$ expands.

\subsection{Description through the matching of geometries}
To start, we review aspects of the thin-wall bubble approximation, 
in order to later state our full numerical results in this simplified language,
and to explain why it is not adequate for the problem at hand.
An approximate description of the evolving bubble can be obtained by assuming the metric
\beq
 ds^2=\left\{\begin{array}{ll}
 \displaystyle
 -A_{\rm in}(r)\,  d t_{\rm in}^2+ \frac{ dr^2}{A_{\rm in}(r)}+r^2  d\Omega^2 & 
 \hbox{for $ r< R$ with  $\displaystyle A_{\rm in}=1+ \frac{r^2}{\ell_{\rm in}^2} -\frac{2 G M_{\rm in}}{r}$}\\[1ex] 
 \displaystyle
 -A_{\rm out}(r)\,  dt^2_{\rm out}+\frac{ dr^2}{A_{\rm out}(r)}+r^2  d\Omega^2 &
 \hbox{for  $r > R$ with 
$ \displaystyle A_{\rm out}=1- \frac{r^2}{\ell_{\rm out}^2}-\frac{2GM_{\rm out}}{r}$}
 \end{array}\right.
 \label{eq:SAdS-SdS} \eeq
where $d\Omega^2 = d\theta^2 + \sin^2\theta\, d\varphi^2$. 
A thin wall located at $r=R(t)$ separates
Schwarzschild-de Sitter (SdS) outside with $1/\ell_{\rm out}^2=\HI  ^2= V_{\rm out}\, 8\pi G /3$,
from Schwarzschild-anti de Sitter (SAdS) inside with
$1/\ell^2_{\rm in}=-V_{\rm in}\, 8\pi G /3$, where $V_{\rm out}>0$ and $V_{\rm in}<0$ are the energy densities.
Similar solutions exist if $V_{\rm in}$ and $V_{\rm out}$ have the same sign.
The solution also contains a central black hole with mass $M_{\rm in}$.
The metric transverse to $r$ is  continuous on the wall, 
while the function $A(r)$ has a discontinuity
proportional to the wall surface tension $\sigma$, as dictated by the Israel matching condition~\cite{Israel:1966rt,Blau:1986cw}
\beq\label{eq:Israel}
\sqrt{A_{\rm in}+\dot{R}^2} - \sqrt{A_{\rm out}+\dot{R}^2}=4\pi G\sigma  R.\eeq
Solving for $M_{\rm out}$ results in an intuitive expression:
\beq M _{\rm out}- M_{\rm in}=-\frac{4\pi R^3}{3} \Delta V + 4\pi R^2 \sigma \sqrt{1+\dot R^2 -\frac{2GM_{\rm in}}{R} -\frac{8\pi G}{3}R^2 V_{\rm in}}, 
\label{eq00} \eeq
where $\Delta V = V_{\rm out}-V_{\rm in}+6\pi G \sigma^2$ also includes the gravitational energy.
Solving eq.\eq{Israel} for $\dot R$  gives 
\beq -1-\dot R^2= U \equiv -\frac{(M_{\rm out} - M_{\rm in}+ 4\pi R^3 \Delta V/3)^2}{(4\pi R^2\sigma )^2} -\frac{2GM_{\rm in}}{R} - \frac{8\pi G}{3}  V_{\rm in} R^2,
\label{evpot}\eeq
a form useful for studying the motion of the wall, 
as it is formally similar to the conservation of `energy' for a point in a `potential' $U$
that encodes the general-relativistic effects.

The solutions of eq.~(\ref{evpot}) for $M_{\rm in}=0$ were discussed in detail 
in \cite{Higgstory}, and the generalization for $M_{\rm in}\not= 0$ was given in 
the appendices of~\cite{1606.04018}. A general feature is the existence of solutions
that describe expanding bubbles. The expansion is energetically favoured for large bubbles,
when the interval volume with negative energy density dominates the total energy budget. 
For example, setting $M_{\rm in}=0$ (no black hole inside) and ignoring the last term of eq.~(\ref{evpot}) (i.e.\ in the limit $G\to 0$), 
the sign of $\partial U/\partial R$ indicates 
that a thin-wall bubble initially at rest expands if $R> 3\sigma/\Delta V$, as is also
expected from the initial expression for constant $M_{\rm out}$.%

However, there are two approximations that must be relaxed when developing 
a formalism that could
approximate an evolving field in the interior. They concern the parameters  $V_{\rm in}$
and  $M_{\rm in}$, which must be allowed to evolve if they are to approximate a time-dependent
field configuration. In the rest of this section we discuss the possible implications,
before studying the full problem in the following section.

\subsection{Interior with a continuous mass distribution}\label{Bondi}
The first step towards a more general description is the replacement of the 
mass parameter $M_{\rm in}$ of the SAdS metric with a continuous mass distribution. 
An analytic description is very difficult for a general equation of state. However, 
a simple analytic solution exists if  $M_{\rm in}$ is attributed to a pressure-less
component. We consider this case as a toy model that illustrates how a result qualitatively
different from~\cite{Riotto} can arise if we 
eliminate the rigidity assumption that links the evolution of the bubble surface
to that of the interior.

We employ the  thin-wall approximation and
describe the space inside the bubble through the Tolman-Bondi (TB) metric \cite{Tolman:1934za,Bondi:1947fta}
in the presence of a negative cosmological constant:
\be
 ds^2=- d \tit^2+ \frac{B'^2(\tit,\rt) }{1+f(\rt)}d\rt^2+B^2(\tit,\rt)  d\Omega^2.
\label{ltb-metric} \ee
The function $B(\tit,\rt)$ gives the location of the shell with comoving coordinate $\rt$
as a function of time. It satisfies 
\begin{eqnarray}
\dot{B}^2&=&\frac{1}{4\pi\bp^2}\frac{M(\rt)}{B(\tit,\rt)}+f(\rt) +\frac{V_{\rm in}}{3\bP^2}B^2(\tit,\rt),
\label{ltb1} \\
M'(\rt)&=&4\pi  B^2(\tit,\rt)\,\rho(\tit,\rt)\,B'(\tit,\rt).
\label{ltb2}
\end{eqnarray}
The function $M(\rt)$ gives the (conserved) integrated mass of the fluid, 
up to the shell with coordinate $\rt$.
% located at  $B(\tit,\rt)$. 
The function $f(\rt)$ can be viewed as a generalized 
spatial-curvature term; it will play no role.
The coordinate patch covers the part of the space for 
which the right-hand side of eq.~(\ref{ltb1}) is positive.

The FRW metric, for a space 
containing a homogeneous pressure-less
fluid, is obtained for 
$B(\tit,\rt)=a(\tit)\rt$, 
$\rho(\tit)={\rho_0}/{a^3(\tit)}$, 
$f(\rt)=k \rt^2$, with $k=0,\pm 1$.
The homogeneous case provides 
intuition about the nature of the evolution in the interior of the bubble.
The resulting Friedmann equation reads
\be
\left( \frac{{\dot{a}}}{a} \right)^2= \frac{1}{3\bP^2}
\left(\frac{\rho_0}{a^3}+V_{\rm in} \right) +\frac{k}{a^2}.
\label{friedmann} \ee
It is apparent that an initially expanding spacetime will stop expanding when $a(\tit)$
reaches a value such that the right-hand side of the above equation vanishes. It will subsequently
collapse to a singularity within a finite time $\tau$. Near the singularity 
we have $a(\tit)\sim (\tau-\tit)^{2/3}$ and the energy density of the pressure-less 
fluid gives the dominant contribution, with a time dependence 
$\rho(\tit) \sim (\tau-\tit)^{-2}$. This is analogous (even though the
exponents of the singular terms differ) to the behaviour deduced in \cite{Riotto}, where
the assumption of homogeneity was also made.

\begin{figure}[t]
$$\includegraphics[width=0.45\textwidth]{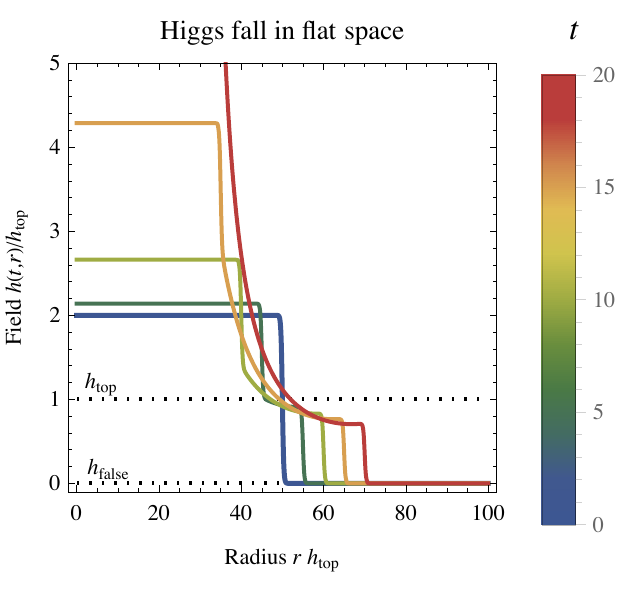}
\qquad \raisebox{0.03\textwidth}{\includegraphics[width=0.4\textwidth]{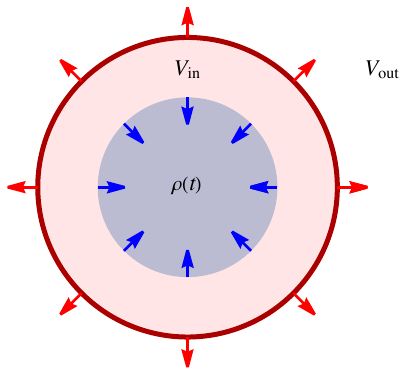}}
$$
\caption{\it {\bfseries Left:} Time evolution in flat space: even assuming an initially thin-wall Higgs profile, its wall expands on both sides,
and the thin-wall approximation breaks down.
We here assumed the SM Higgs potential of eq.\eq{VhSM}
and an initial field configuration with with $\dot h =0$ and $h=2 h_{\rm top}$ for $ r< 10/h_{\rm top}$
and $h=0$ outside. 
The spatial profile of the potential energy $V(h(t,r))$ resembles a sink-hole.
{\bfseries Right:} The toy model discussed in section~\ref{Bondi} replaces
the thin-wall configuration with rigid interior by an onion-like structure,
showing that the collapse of the interior does not imply the collapse of the exterior.
% (also known as `portal to hell').
%\AS{as qualitatively illustrated in the right-handed plot.
\label{fig:HiggsFallFlatSpace}}
\end{figure}

\smallskip

If the matter is concentrated within a radius $r_{\rm mat}$, so that $\rho(\tit,\rt)=0$ and
$M(\rt)=M(r_{\rm mat})$ for $\rt>r_{\rm mat}$, the geometry outside this region is the 
standard SAdS geometry. 
This can be made explicit by matching the interior 
metric in eq.\eq{SAdS-SdS} with the metric in eq.~(\ref{ltb-metric}) at some $r>r_{\rm mat}$ through the 
Israel matching conditions.  
A smooth matching at $r=B(\tit,\rt)$  requires $M_{\rm in}=M(r_{\rm mat})$.
In this way the configuration illustrated in fig.\fig{HiggsFallFlatSpace}b appears consisting of: a) a central inhomogeneous region 
up to $r_{\rm mat}$, b) a shell of SAdS space up to the bubble surface, c) the bubble surface with
constant tension, d) the SdS exterior of
the bubble. The evolution of the surface is described by the formalism of the previous
subsection.
It is clear then that, in this case, the evolution of the bubble surface is 
blind to the actual mass distribution and is affected only by the total mass.  
The analysis is identical to~\cite{1606.04018} and shows that the gravitational 
collapse, by itself,
does not prevent the bubbles from expanding.

The situation is more complicated if the matter distribution extends all the way to
the bubble surface. 
%\AS{Are you considering matter surrounded by a thin wall?}
However, a particular case is illuminating.
We assume that the nonrelativistic fluid covers the whole interior 
of an expanding bubble at the 
time at which $\dot{B}(\tit_m,r_{\rm mat})=\dot{a}(\tit_m)r_{\rm mat}=0$.
This is the time of the maximal expansion of the interior. 
At later times the interior starts contracting, 
so that the shell with comoving coordinate $r_{\rm mat}$
moves inside the bubble. The assumption of a constant bubble tension implies that 
there is no source of matter on the
bubble surface. It is, therefore, expected that a gap will appear between the 
collapsing matter and the bubble radius. Within this part of space, the metric is of the
SAdS form, with $M_{\rm in}=M(r_{\rm mat})$.
The crucial point is that 
the evolution of the bubble surface remains unaffected by all that is happening in its
interior, so that it can keep expanding.
Matter can collapse forming a black hole, while the wall expands outside the horizon.

\smallskip

Allowing for radial inhomogeneities or different initial conditions prohibits 
a precise description. However, several features of the evolution can be 
deduced intuitively. The effect of the interior on the bubble evolution is 
parametrised by the mass function $M(r_{\rm mat})$ at the location of the surface.
In this sense, the relative expansion or contraction of various regions in the
interior plays a secondary role.
In general, the value of $r_{\rm mat}$ is time-dependent. 
If it increases with time, matter may
either accumulate on the bubble surface, or move into the exterior. 
The first possibility would violate our assumption of a constant wall tension.
The second possibility would result in matter leaking into the region dominated
by a positive cosmological constant, where it would be diluted by expansion.
This would continue until the interior starts collapsing.
These observations indicate that the approximate description of the interior by 
a SAdS metric with a mass parameter corresponding to the total mass within
the bubble gives a reasonable description of the dynamics. The 
general conclusion is that the bubble evolution
is not tied to the collapse of the interior.

Even though this toy model provides an intuitive understanding of the competing 
features of the evolution, it is not adequate for the description of a fully dynamical
field. We turn to this problem next.

\begin{figure}[t]
$$%\includegraphics[width=0.45\textwidth]{figs/finfoutsampledSMneg}\qquad
\includegraphics[width=0.45\textwidth]{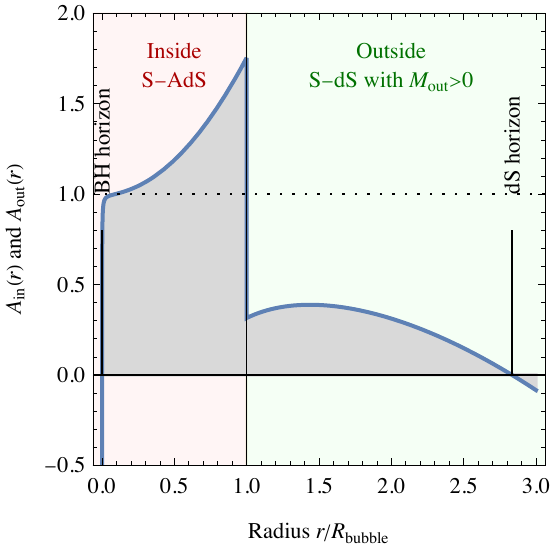}\qquad
\includegraphics[width=0.45\textwidth]{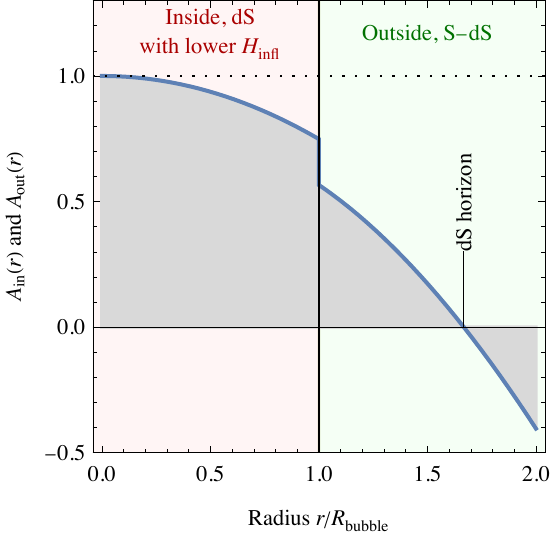}$$
\caption{\it {\bfseries Left:} 
Typical form of $A(r)$ in the thin-wall approximation for the AdS bubble 
inside dS considered in~\cite{Higgstory}.
The discontinuity is proportional to the wall tension $\sigma$.
Here $\sigma$ is assumed to be large enough for $M_{\rm out}>0$.
%In the right plot $\sigma$ is large enough that $M_{\rm out}>0$.
%In the left plot $M_{\rm out}<0$ dominated by the large inside volume,
%and not compensated by the small tension
%(in such a case no critical bubble exists: bubbles either expand or contract).
{\bfseries Right:} Typical form of the initial condition assumed
in~\cite{Riotto} and in the present paper.
Smaller values of $A$ loosely mean that time runs more slowly. 
\label{fig:finfoutsample}}
\end{figure}

\subsection{Scalar at its minima and applicability of the thin wall approximation}
The `Higgstory' study~\cite{Higgstory} assumed as initial condition a thin wall bubble
with the scalar $h$ near to its minima both inside and outside:
$h_{\rm out} \approx h_{\rm false}$ and $h_{\rm in}\approx h_{\rm min}$.
The resulting geometry
is exemplified in fig.\fig{finfoutsample}a in the language of eq.\eq{SAdS-SdS}
and the above formalism implies that such thin-wall bubble can expand.
%\footnote{Furthermore $V\le 0$ has a maximum $V=0$ at $R^3 = 3M/4\pi \Delta V$.
%Writing $M$ in terms of $R_0$ (the radius at which $\dot R=0$) this becomes $R^3 = R_0^2 (R_0  - 3 \sigma/\Delta V)$,
%showing again the critical value $R_0 =2\sigma/\Delta V$.
%So this is just a more indirect way of computing.}
%\footnote{For $G\neq 0$ a thin-wall bubble initially at rest expands if larger than
%\beq R > \frac{2\sigma/\Delta V}{\sqrt{1/2+\sqrt{1/4+\epsilon/3}+\epsilon}}\simeq\left\{\begin{array}{ll}
%(1-2\epsilon/3) \sfrac{2\sigma}{\Delta V} & \epsilon \ll 1\\
%  1/\sqrt{4\pi  G V_{\rm in}} = \sqrt{2/3}\ell_{\rm in}& \epsilon \gg 1\end{array}\right.
%\qquad
% \epsilon =\frac{16\pi GV_{\rm in}  \sigma^2}{\Delta V^2}
%\eeq
%So AdS stops the expansion of bubbles smaller than the AdS radius.
%Deep AdS stops less.}

In this case the thin-wall approximation is adequate because in the subsequent evolution $h$ remains close to its minima,
and even an initially thick wall becomes thin thanks to its expansion.
Indeed the wall tension $\sigma$ can be approximated assuming that in a small range $\Delta r$ the scalar field
varies by $\Delta h = h_{\rm in}-h_{\rm false}$: 
\beq \label{eq:sigma}
\sigma \approx \int dr \bigg[\frac12 \bigg(\frac{\partial h}{\partial r}\bigg)^2+V(h) - V(h_{\rm in})\bigg]
\sim \frac{\Delta h^2}{\Delta r} + \Delta r  \Delta V \circa{>} \Delta h\sqrt{\Delta V}\eeq
minimised for a bubble thickness $\Delta r \sim  \Delta h/\sqrt{\Delta V}$
comparable to the minimal radius such that the bubble expands.
In this case $h_{\rm in}=h_{\rm min}$ remains fixed during the evolution,
so $\sigma$ and the shape of the thin wall remain fixed
(as assumed in eq.\eq{Israel}) and only the location $R$ of the wall can evolve.
For a quartic potential $\sigma \approx \sqrt{|\lambda|} \Delta h^3$.

\subsection{Scalar not at its minima and inapplicability of the thin wall approximation}
The recent study~\cite{Riotto} considers a different initial condition at an earlier time:
the Higgs expectation value inside the bubble, $h_{\rm in} $, has not yet reached the minimum of its potential.
Rather, the field inside the bubble is assumed to have a value just past the potential barrier,
$h_{\rm in} \circa{>} h_{\rm top}$,
with the potential inside taking a value much above $V_{\rm min}$, $V_{\rm in}\circa{<} V_{\rm false}$.
The resulting cosmology is inflationary on either side of the bubble surface, but
with a lower expansion rate inside, as illustrated in fig.\fig{finfoutsample}b.

Since $h_{\rm in}(t)$ falls down the potential towards $h_{\rm min}$, 
the gap in field values $\Delta h = h_{\rm in}- h_{\rm out}$ grows with time.
Therefore, the wall tension $\sigma$, estimated in eq.\eq{sigma}, 
cannot remain constant in time, unlike what was assumed in~\cite{Riotto}
(leading to the claim that the bubble disappears).
It is not even possible to improve the thin-wall approximation 
by allowing for a time-dependent  wall tension  $\sigma(t)$ 
and adding a time-dependent wall pressure.
Indeed, if the wall is not surrounded by regions where $h$ is at its minimum,
the wall has more degrees of freedom than the one corresponding to radial displacement.
Even assuming an initially thin wall, the field can now fall down the potential on 
{\em both} sides,
getting stretched at roughly the speed of light by an amount $\Delta r \sim t$.
As a simple example of this, fig.\fig{HiggsFallFlatSpace} displays the numerical 
solution to the classical Higgs equation of motion
$\ddot h - h'' -2 h'/r + V'=0$ 
(in the usual notation $\ddot h = \partial^2 h/\partial t^2$, $h' 
= \partial h/\partial r$, $V'=\partial V/\partial h$), assuming
a spherical Higgs field configuration $h(t,r)$
in the SM potential that is approximated through an effective running coupling as
\beq\label{eq:VhSM}
V \approx \lambda(h)\frac{h^4}{4} ,\qquad
\lambda(h)=-b \ln \frac{h^2}{e^{1/2}h_{\rm top}^2}\qquad\hbox{with}\qquad  b \approx \frac{0.15}{(4\pi)^2} .
\eeq
The electroweak vacuum is located at $h_{\rm false}\approx 0$, while the true minimum is so 
deep that its existence plays no significant role.
Fig.\fig{HiggsFallFlatSpace} shows that an initial thin-wall field profile is not maintained in the following time evolution.
The potential energy of the system resembles a sinkhole
rather than a wall.
As the fall of $h_{\rm in}$ inside proceeds, the potential barrier in $V$ becomes 
negligible and the scalar field profile approaches a different general form:
deeper where it had more time to fall.
This kind of time evolution is expected to be general, since it only depends on local physics around the initially thin wall.
The wall evolution violates the assumption, used to justify the thin-wall approximation, 
that the domain wall 
settles into an equilibrium configuration (see e.g.~section III of~\cite{Blau:1986cw}).

\smallskip

After a time of order $\tau=1/h_0 \sqrt{|\lambda|/2}$\footnote{Indeed, 
the solution to $\ddot h+ \lambda h^3=0$ is $h(t)= h_0/(1- t/\tau)$
assuming a constant $\lambda<0$ and an $r$-independent initial field value $h_0$ at $t=0$.}
the Higgs fall reaches the deep minimum and bounces back,
or hits a singularity if the potential is approximated as unbounded from below.

We emphasize that the subsequent evolution is computable even after the formation of the singularity, 
in the region outside its light-cone and thereby not causally affected by the singularity. 
The generic outcome is that the bubble keeps expanding at the speed of light.
The simplest computable example of this phenomenon is a potential
$V = \lambda h^4/4$ with $\lambda<0$ in flat space,
as the classical equation $ \ddot h - h'' + 2 h'/r + V'=0$
admits the Fubini solution
\beq h(t,r) = \frac{h_0}{1+(r^2-t^2)/r_0^2}\qquad  \hbox{with} \qquad h_0= \frac{1}{r_0}\sqrt{-\frac{8}{\lambda}}.\label{eq:Fubini}\eeq
This solution describes a thick wall that keeps expanding at the speed of light after that the singularity
appears at a time $t=r_0$.
One way of obtaining numerical solutions consists in `regularising' the divergence by 
adding to the potential a deep true vacuum,
such that the region affected by the singularity is replaced by the scalar field 
oscillating around the minimum,
while the causally disconnected region at large value of $r$ is not affected.
Overall, the system is analogous to a sinkhole in the ground 
that keeps expanding while the parts near the surface keep sliding down, 
irrespectively of the fact that its central region might be infinitely deep.

\medskip

The above discussion and the numerical simulation in fig.\fig{HiggsFallFlatSpace} 
have neglected gravity.
According to~\cite{Riotto} gravity adds one key new effect: a thin-wall bubble disappears 
in a gravitational collapse,
getting fully hidden behind a horizon.
Since we have argued that the system cannot be approximated by a thin wall, we expect that 
the scalar field dynamics, together with gravitational dynamics,
will remove the inner part of the system,
but the outer part that unavoidably develops (even around an initially thin wall) 
will survive and keep expanding.

Even if only one bubble finds a way to expand and engulf the Universe, 
the usual bounds (see~\cite{Higgstory} and subsequent papers) apply.
This makes it possible to settle the issue through an example.
We thereby proceed to solve the equations of the gravity-Higgs system 
 in section~\ref{sinkhole} in order to
produce numerical examples, as well as a particular analytic solution.

%\subsection{Onion generalization of the thin wall}
%\AS{WRITE little onion. We could put it here, as an example of a 3-stage toy system that extends the thin wall?}
%Since there is no deep fall, this kind of solution is qualitatively similar to what analytically described by~\cite{Bondi:1947fta}:
%an onion with a thin wall and non-relativistic matter inside
%(the non-relativistic assumption is needed to obtain an analytic solution)
%as shown in Appendix~\ref{Bondi} matter falls forming a black hole, while the wall can remain outside an expand.

\section{Gravitational Higgs sinkhole}\label{sinkhole}

In section~\ref{analytic} we present a special analytic solution that 
describes a scalar rolling down its potential, taking gravity into account.
We next obtain numerical solutions in generic situations.
To cross-check our results we use two different ansatze for a metric with spherical symmetry:
reparametrization invariance is used in section~\ref{FRW} to make the time dependence
explicit, resulting in a FRW-like metric; 
while in section~\ref{static} the radial dependence is made explicit,
resulting in a Schwarzschild-like metric.

Before presenting the solutions, 
we recall a tool that will help the interpretion of singularities appearing 
in time-dependent spherically symmetric geometries.
An apparent horizon
%~\cite{Abbott:1985kr,gr-qc/9303006,0807.3797,Riotto} 
is defined as the boundary of the region of 
trapped surfaces. This boundary is a surface on
which the expansion of outgoing null geodesics vanishes. 
In order to determine the presence of such a horizon,
we examine the expansion of radial null geodesics.
The two sets of geodesics, generically denoted by $r_\pm(t)$, 
can be determined from the metric.
In flat space $dr_\pm/dt = \pm 1$, so that 
the solutions $r_+(t)$ and
$r_-(t)$ clearly correspond to outgoing and ingoing null geodesics, 
respectively. However, in non-trivial geometries, and especially in the vicinity of
horizons, a more careful analysis is necessary in order to 
determine their nature. (Despite this, 
we always refer to the geodesics $r_\pm(t)$ as out/ingoing, for simplicity). 
%This characterization  is strictly consistent in the asymptotic regions outside the horizons. On the other hand, 
For a spherically symmetric geometry, the true nature of the geodesics 
becomes clear if we consider that they define surfaces of areal radii 
$R_\pm(t,r_\pm(t))$. A truly outgoing geodesic results in the growth of the 
area of such a surface, while an ingoing geodesic results in the
reduction of the area. On an apparent horizon, the rate of change of
the area vanishes. The product
\be
\Theta= \frac{dR_+}{dt}  \frac{dR_-}{dt} \label{eq:Thetaspherical} \ee
is a convenient quantity in order to search for horizon as it is
independent of the normalization of the vectors that 
define the null hyper-surfaces. In flat space $dr_\pm/dt = \pm 1$ so $\Theta=-1$. An 
apparent horizon would appear at the point where $\Theta$ vanishes before changing sign.

\begin{figure}[t]
$$\includegraphics[width=0.45\textwidth]{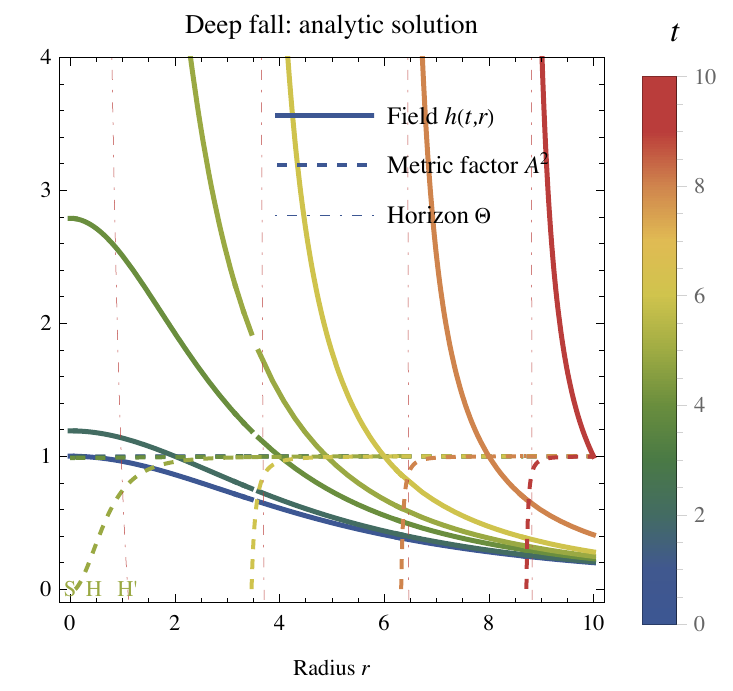}\qquad \includegraphics[width=0.45\textwidth]{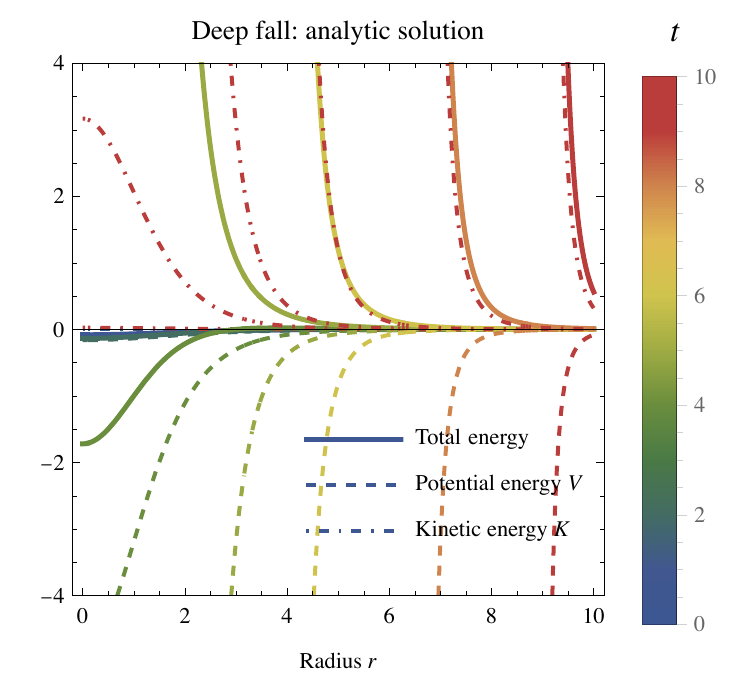}$$
\caption{\it Evolution in Einstein gravity of the scalar bubble for the potential of
eq.\eq{FubGrav}.
{\bfseries Left}: the field $h(t,r)$, the scale factor $A(t,r)$, and the criterion for 
horizon formation $\Theta(t,r)$.
An apparent horizon first appears at the point denoted as {\rm H}. 
Later a singularity appears inside the horizon at the point denoted as {\rm S}, while the horizon reaches ${\rm H}'$.
{\bfseries Right}: the energy densities. 
We assumed $h_0=\bP/10 $, $r_0=5/h_0$ and plotted in units in which $h_0=1$.
\label{fig:DeepFallAnalytic}}
\end{figure}

\subsection{Special analytic solution}\label{analytic}
We start by presenting a special analytic solution for an expanding scalar bubble in general relativity.
It is obtained starting from the analytic Fubini solution of eq.\eq{Fubini}, valid in the
absence of gravity
for a purely quartic scalar potential.
Since the potential is scale-invariant, eq.\eq{Fubini} remains a solution even 
in the presence of gravity, provided that the scalar $h$ has a 
conformal coupling to gravity.
Starting from this solution and rewriting the system in the Einstein frame in terms of 
a canonically normalized scalar $h$, 
one obtains the potential 
\beq \label{eq:FubGrav}
V = 9\lambda \bP^4 \sinh^4\frac{h}{\sqrt{6}\bP}.
\eeq
The Einstein and scalar-field equations
have the solution 
\beq \label{eq:metriconf}
h=\sqrt{6}\bP \atanh\frac{h_0/\sqrt{6}\bP}{1 + (r^2 - t^2)/r_0^2},
\qquad 
g_{\mu\nu} =A^2(t,r) \eta_{\mu\nu}
= \left[1-\frac{h_0^2/6\bp^2}{(1+(r^2-t^2)/r_0^2)^2}\right] \eta_{\mu\nu},
\eeq
with a conformally-flat metric, arbitrary $h_0$, and
$r_0 =  \sqrt{-8/\lambda}/h_0$.

This solution is visualised in fig.\fig{DeepFallAnalytic}.
The singularity with $h\to\infty$, 
$A\to0$ and infinite curvature first develops at $r=0$ 
at a time $t_{\rm s}=r_0 [1-h_0/\sqrt{6}\bP]^{1/2}$.
At later times the location of the singularity moves to finite values of $r$.
However, $A$ vanishes at the same $r$, so that physical distances, such as the 
areal distance $Ar$, also vanish.

As discussed above, an apparent horizon is present 
if $\Theta$, defined in eq.\eq{Thetaspherical}, vanishes.
In this example, 
outgoing/ingoing geodesics for the metric of eq.\eq{metriconf} 
satisfy $dr_\pm /dt=\pm 1$, while the areal radius is $R=Ar$. This gives
\be
\Theta=(\dot{A} r)^2-(A+A' r)^2.
\label{thetaconformal} \ee
As illustrated in fig.\fig{DeepFallAnalytic}, the singularity is always 
surrounded by the apparent horizon.
Overall, the analytic solution is qualitatively similar to the generic 
cases computed via numerical tools and presented in the following subsections.
It clearly
shows that the bubble keeps expanding at the speed of light, 
even after the formation of the singularity,
with the total energy density $V + \dot h^2/2A^2 + h^{\prime 2}/2A^2$  
becoming positive.\footnote{The analytic continuation of our solution into the region
behind the singularity can be interpreted as the creation of a Universe induced by scalar inflation.}
This provides a simple counter-example to the possibility of a general gravitational 
mechanism that prevents bubble expansion.

\subsection{Metric in time-friendly FRW form}\label{FRW}
A spherical non-homogeneous system can be described by the time-friendly FRW-like metric~\cite{Bondi:1947fta}
\beq ds^2 = -dt^2 + [a^2(t,r) dr^2 + b^2(t,r) r^2 d\Omega^2]\eeq
so that $t$ is a simple time parameter and $br$ is the areal distance.
%Classical equations can be computed by expanding the action in terms of $h,a,b$ and
%imposing vanishing functional derivatives.
The resulting field equations contain second  time derivatives for all fields $h,a,b$:
\begin{eqnsystem}{sys:hab}
\ddot h +\dot h \bigg(\frac{\dot{a}}{a} + 2\frac{\dot b}{b}\bigg)
 &=& - V' + \frac{h''}{a^2} + \frac{h'}{a^2} \bigg(\frac{2}{r} - \frac{a'}{a}+2 \frac{b'}{b}\bigg), \\
 \frac{\ddot b}{b}+\frac{\dot b^2}{2 b^2} &=& 4\pi G\bigg( V - \frac{\dot h^2}{2} - \frac{h^{\prime 2}}{2a^2}\bigg)
-\frac{1/b^2-1/a^2}{2 r^2} +\frac{b'(2b+rb')}{2a^2 b^2 r},
 \\
\frac{\ddot{a}}{a}  +\frac{\dot{a} \dot{b}}{a b}-\frac{\dot{b}^2}{2 b^2}
   &=&  4\pi G \bigg(V-\frac{\dot{h}^2}{2} +3 \frac{ h^{\prime 2}}{2 a^2}\bigg) +\frac{1/b^2-1/a^2}{2 r^2} +\\
   &&+   \frac{1}{a^2}\frac{b''}{b} + \frac{ab'(2b-rb')-2ba'(b+rb')}{2 a^2 b^2 r} .\nonumber
  \end{eqnsystem}
We numerically solve the equations for a potential of the form of eq.\eq{V}.
The barrier is at $h_{\rm top}=m/\sqrt{|\lambda|}$ in the relevant bottom-less limit $\Lambda=\infty$.
We consider the initial condition $\dot h=0$, $a=b=1$, $\dot a=\HI \approx \dot b$ at $t=0$.
The initial field profile $h(0,r)=h_0(r)$ is assumed to be
\beq \label{eq:h0}
h_0 (r) = \left\{ \begin{array}{ll}
h_0/(1+r^2/r_0^2) & \hbox{thick wall,}\\
h_0 [1-\tanh(c (1-r/r_0))]/2& \hbox{thin wall for $c \gg 1$.}
\end{array}\right.\eeq
The parameters $h_0$ and $r_0$ of the
initial configuration $h_0(r)$ must be chosen such  $h(t,r)$ evolves towards its true minimum.
\begin{itemize}
\item We choose $h_0 \circa{>} h_{\rm top}$ beyond the potential barrier
but still in the region where $  V(h_0) >0$, such that one has
inflationary de Sitter space with Hubble constant $\HI$ at $r \gg r_0$, 
and de Sitter with lower Hubble constant at $r \circa{<} r_0$.

\item Furthermore, as discussed around eq.\eq{sigma},
in the thin-wall limit a bubble expands for $r_0 \circa{>}\Delta h/\sqrt{\Delta V} \sim 1/h_0\sqrt{|\lambda|}$,
where $\Delta h$ and $\Delta V$ are the field and potential variations along the bubble.
Beyond the thin-wall limit no such simple criterion for expansion is known,
but the thin-wall criterion remains qualitatively correct.
\end{itemize}
Motivated by inflationary dynamics, we consider a Hubble scale comparable to the size of the bubble.
Then $\Delta V \ll V$ for $h_{\rm top} \ll M_{\rm Pl}$. % i.e. $G \ll 1/h_{\rm top}^2$.
More precisely, we display
the result for the following numerical example\footnote{Avoiding large or small numbers helps the 
computation and the visualization of the numerical solution.
Similar results are obtained in the SM-like case, 
where the key collapse dynamics happens within a small fraction of the time range.}
\beq\label{eq:esempio} \lambda=-1 ,\qquad \HI=m = h_{\rm top} = M_{\rm Pl}/10,\qquad r_0 = 5/m,\qquad h_{\rm 0}=2h_{\rm top}.\eeq
Another important scale of the problem is the time needed for the deep fall of $h$.
Taking gravity into account, it is estimated as
\beq\label{eq:tau} \tau=\frac{3\HI}{2 |\lambda| h_0^2} \eeq
by considering a homogeneous Higgs field in a fixed inflationary background: 
neglecting $\ddot h$ in its equation $\ddot h + 3 \HI  \dot{h}+ \lambda h^3=0$,
one finds the solution $h(t) \approx h_0/\sqrt{1-t/\tau}$~\cite{Riotto}.

\begin{figure}[t]
$$\includegraphics[width=0.45\textwidth]{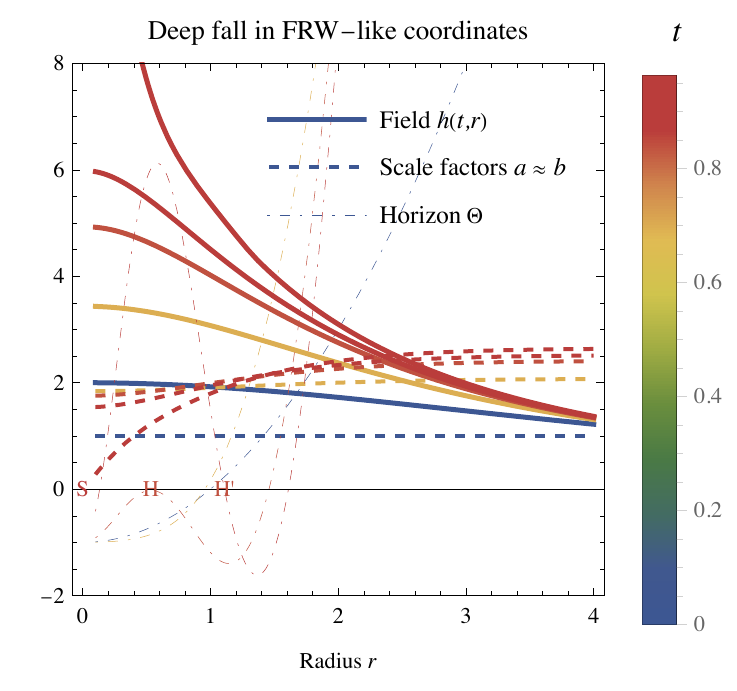}\qquad \includegraphics[width=0.45\textwidth]{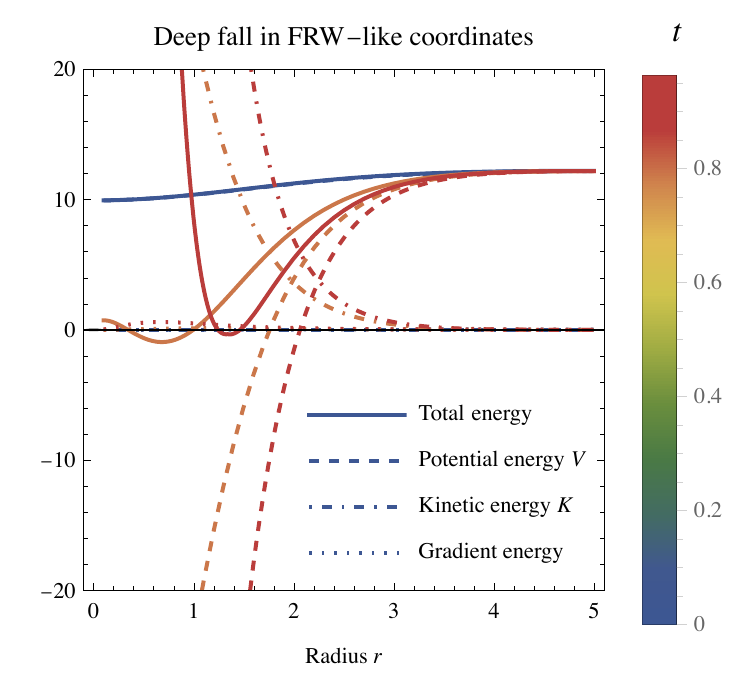}$$
\caption{\it Numerical evolution in Einstein gravity of the scalar bubble 
beyond the top of the potential described around eq.\eq{esempio}.
We work in units in which $h_{\rm top}=1$.
{\bfseries Left}: the field $h(t,r)$, the scale factors $a(t,r)\approx b(t,r)$; 
the criterion for horizon formation $\Theta(t,r)$.
A black-hole horizon first appears at the point denoted as {\rm H}. 
Later a singularity appears inside the horizon at the point denoted as {\rm S}, while the horizon reaches ${\rm H}'$.
{\bfseries Right}: the energy densities. 
\label{fig:DeepFallFRWha}}
\end{figure}

The quantity $\Theta=0$ of eq.\eq{Thetaspherical}, used to deduce the
formation of an apparent horizon,
becomes
\be
\Theta \equiv b^2 r^2 \left[\frac{\dot{b}^2}{b^2}
-\frac{1}{a^2} \left(\frac{1}{r}+\frac{b'}{b} \right)^2
 \right]
\label{eq:ThetaFRW} \ee
as the out/ingoing null geodesics
obey $dr_\pm/dt =\pm 1/a(t,r)$, while the areal distance is 
$R_\pm(t)=b(t,r_\pm(t))\,r_\pm(t)$, so that eq.\eq{ThetaFRW} follows from $\dot R_\pm =\dot b r \pm (b+b'r)/a$. 
In the homogeneous limit $a=b =e^{\HI t}$ this reduces to
$\Theta = (r \HI  e^{\HI  t} )^2 - 1$,  reproducing the usual dS horizon at $ar = 1/\HI $.
%(in-going rays get inflated).

%Fig.\fig{} shows that an extra horizon appears at radius $r \approx 0.5$ and moves out.

\subsubsection*{Thick-wall numerical solution}
If the true minimum $V(h_{\rm min})$ of the potential $V(h)$ is not deep, the system evolves in a way qualitatively
similar to the flat case: the field reaches $h_{\rm min}$, bounces and starts oscillating
around the true minimum, while the bubble expands at nearly the speed of light,
following the de Sitter geometry of the outer space-time.

In the opposite limit, with a true minimum  so deep that its presence is irrelevant,
numerical simulations such as the one shown in fig.\fig{DeepFallFRWha}
display the following characteristics.
The scale factors $a\approx b$ initially increase until 
 the total energy density becomes negative around the interior of the bubble
 (due to the fall of the scalar $h$ and to Hubble friction).
Within this region and  at this point in time the scale factors $a\approx b$ start decreasing, 
triggering a run-away accelerated fall of $h$, with energy dominated 
by the kinetic energy (see fig.\fig{DeepFallFRWha}b).
As a result, a singularity in $h$ and in the curvature develops, at the point $r=0$ and at a time $t_{\rm s}\sim \tau$ as in eq.\eq{tau}.
An apparent horizon forms at an earlier time at finite $r$, denoted as `H' in fig.\fig{DeepFallFRWha}a.

In contrast to the thin-wall expectation 
in~\cite{Riotto}, a growing region where the scalar $h$ is mildly
above its potential barrier remains in the region outside the collapse, 
where the scale factors $a\approx b$ keep growing.
 
%At this moment $h(r) > h_{\rm top}$ even in outer regions  keeps expanding. 

Regions inside the apparent horizon do not affect the exterior, so that
we can keep computing after the singularity develops by dropping regions in its
immediate vicinity, confirming that
the $h(t,r)$ bubble keeps growing, similarly to 
the analytic solution of section~\ref{analytic}.
We also point out that the scale factors $a\approx b$ vanish at the location 
of the singularity, indicating that the areal distance for this point also vanishes.

The numerical solution shows that the thin-wall  claim of~\cite{Riotto}  cannot be fully general.
As one bubble that expands is enough to imply bounds, 
having demonstrated its existence settles the wider issue.
However, in order to obtain a broader perspective
we next proceed to examine wall profiles that are initially thin.

%qualitatively different from a thin-wall
%(if the fall is fast, roughly each point falls as much as it can),
%and remains large and above the barrier
%(while it would vanish in the thin-wall approximation of~\cite{Riotto}).

\begin{figure}[t]
$$\includegraphics[width=0.45\textwidth]{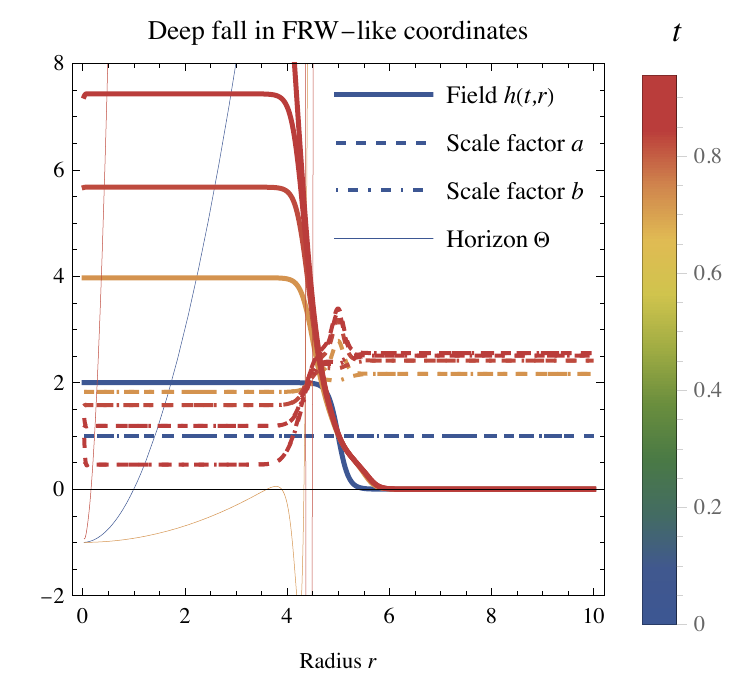}\qquad \includegraphics[width=0.45\textwidth]{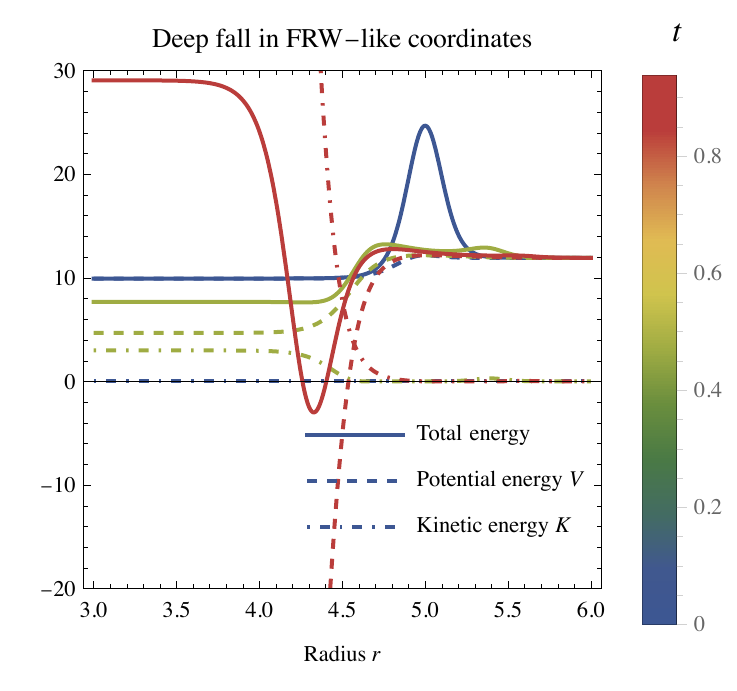}$$
\caption{\it As in fig.\fig{DeepFallFRWha}, but considering an initially thin wall. 
\label{fig:DeepFallFRWthin}}
\end{figure}

\subsubsection*{Thin-wall numerical solution}
In order to highlight the difference with respect to the thin-wall claim of~\cite{Riotto},
we next evolve a special initial configuration with an initially thin wall, 
namely the profile in the lower row of eq.\eq{h0}, 
such that $h_0(r)$  is piece-wise nearly constant:
inside at $r< r_0$, and outside at $r>r_0$.

We again consider a deep fall of the scalar field (otherwise the scalar soon reaches its true vacuum and bounces back, leading to
the usual expanding bubble of true vacuum).
The numerical solution in fig.\fig{DeepFallFRWthin} shows that, as expected, the thin wall approximation breaks down.
%Nevertheless, this special case also helps to partially understand the solution.
At the beginning the scale factors $a, b$ grow while remaining piece-wise constant, being 
smaller inside that outside the bubble.
Their equality $a\approx b$ is only violated around the bubble surface
%(thanks to the vanishing of the only $r$-dependent term not related to spatial gradients in %eq.~(\ref{sys:hab}),
%proportional to $1/a^2-1/b^2 \approx 0$).
%The horizon-tagging function $\Theta(t,r)$ is constant, 
%$\Theta\simeq - 1$, at the moment when the inside expansion stops, $\dot a=\dot b=0$. 
An apparent horizon forms around the formerly-thin wall at $r \approx r_0$
as soon as the collapse starts in the interior.
Indeed $\Theta_{\rm in}= r^2 \dot a^2 - a^2$ grows inside, as $\dot a^2$ gets large and and $a$ gets small.
%and $a\approx b$ decrease, 
%Next $a,b$ become (lower inside): $\Theta$ develops a jump at the wall,
%as lower $H$ inside reduces $\Theta$ inside, pushing the dS horizon outwards (if it is inside).
%Next, 
As the outside keeps inflating,
near the wall there is a point where $\dot a=0$: here $\Theta$ remains small and horizons appear.\footnote{
Fig.\fig{DeepFallFRWthin} shows a numerical solution with $\HI r_0 >1$ so de Sitter horizons also play a role.
The same main features apply for $\HI r_0 < 1$ and even with outer Minkowski space, $\HI=0$.}

In the generic thick-wall configuration of fig.\fig{DeepFallFRWha} the FRW scale-factors hit the singularity $a \approx b\approx 0$ at $r=0$.
In the special initially-thin configuration of fig.\fig{DeepFallFRWthin} the scale factors 
$a\approx b$ approach zero at the same time
$t_{\rm s}\approx \tau$ in almost all of the interior.
This makes no difference to an outside observer, as this happens when the wall is no longer thin: 
a growing region with $h$ mildly above the instability critical value
$h_{\rm top}$ and size $\sim \tau $ has developed outside.
We can measure its size in terms of the physical areal distance $b\, r$:
since $b \circa{>} 1$ outside we find 
that the bubble remains big despite the collapse inside.
%\footnote{This footnote clarifies an additional possible doubt, finding that it can be ignored.
%A different picture would be obtained defining a `radial distance' as $dr_{\rm rad}= a \, dr$ such %that $ds^2 = - dr_{\rm rad}^2 + \cdots$.
%Plotted in terms of $r_{\rm rad}$ our solution looks like a mostly-disappearing
%bubble interior, leaving only a tiny BH with Higgs hair.
%But $r_{\rm rad}$ is not the length relevant for an outside observer; 
%e.g.\ $r_{\rm rad}$ gives a non-sense even applied to a Schwarzschild black hole metric.
%The two distances would agree, $r_{\rm rad}=r_{\rm ar}$, if $a = (br)'$.
%In our thin-wall numerical solutions the equality $a = (br)'$ is not true near the wall surface.
%The equality applied in the simplified model analytically solved in section~\ref{Bondi}.}

In a qualitative sense, the crucial difference with respect to the
thin-wall claim of~\cite{Riotto}, according to which the final state only contains a black hole,
is that the black hole lives in the true vacuum rather than in the false vacuum.
It is time-dependent as it accretes 
the energy difference stored in the dynamical scalar field. An outside observer sees an expanding bubble.

%that could invalidate the thin-wall approximation,
%but numerics shows that $a\neq b$  only at the wall. So thin wall is better:
%its only failure is that $h$ develops a tail at $r > R$, .
%Furthermore, numerics shows that the horizon $\Theta$ appears very near to the bubble wall, 
%[and near to the dS horizon? WHY? Coincidence? Or like a contracting sun with constant density: 
%the horizon always appeares around its surface?]

%Even simpler is the case with $\HI =0$, as the wall expands outside at $v=c$ in flat space;
%it expands inside and a horizon appears immediately??
%The fall is so fast that having inflation does not matter too much: one could simplify the problem, but we anyhow keep it.
% Bondi can analytically solve it for zero pressure?

\subsection{Metric in space-friendly static form}\label{static}
As a check, we next solve again the same physical problem replacing the time-friendly FRW coordinate system
with a space-friendly static coordinate choice.
The metric is again written in terms of two functions, now called $A$ and $\delta $:
\beq ds^2 =- A(t,r) e^{2\delta(t,r)}\,dt^2 + \left[\frac{dr^2}{A(t,r)} + r^2 d\Omega^2\right].\label{eq:dsAdelta}\eeq
In these coordinates $r$ is the areal distance.   
The classical equations  are
\begin{eqnsystem}{sys:Adelta}
 \ddot{h} - A^2 e^{2\delta} h''&=& \dot{h}\bigg( \frac{\dot A }{A} + \dot \delta\bigg)+h'  A^2 e^{2\delta}\left(\frac{2}{r} + \frac{A'}{A}+\delta   '\right)-
  A e^{2\delta}V' ,\label{eq:eqhAdelta}
  \\
%e^{-2 \delta } \frac{ \ddot{h}}{A^2} -h''&=&e^{-2 \delta } \dot{h}
%\frac{ A \dot{\delta }+\dot{A}}{A^3}+h' \left(\frac{A'}{A}+\delta   '+\frac{2}{r}\right)-\frac{V'}{A}\\
   %
 \delta  '& = &   4 \pi  G  r \left(h^{\prime 2} + e^{-2 \delta }\frac{ \dot{h}^2 }{ A^2}\right),\\
A'+\frac{A-1}{r}& = & 8 \pi  G r \left(-\frac{1}{2} A h^{\prime 2}- e^{-2 \delta }\frac{ \dot{h}^2}{2 A}-V\right).
   \end{eqnsystem}
%The fact that $\delta '>0$ justifies he $e^{2\delta}$ exponential assumed in the metric eq.\eq{dsAdelta} 
%results in $\delta '>0$.
In these static-like coordinates the equations for the metric factors
$A$ and $\delta$ contain no time derivatives
(unlike what happened with FRW-like coordinates).
So the whole system is equivalent to one integro-differential equation for $h$.
Since $\delta$ is a time-dilation factor, it becomes irrelevant in the static limit: the equation for  $\delta$ separates from the others,
as one can see by substituting $\delta'\ge 0$ and $A'$ in the equation for $h$ obtaining
\beq \frac{1}{Ae^{2\delta}}\bigg[  \ddot h -e^{-2\delta} \dot h \bigg( \frac{\dot A }{A} + \dot \delta\bigg)\bigg]
=  A  h''+h' \left(\frac{1+A}{r} - 8\pi G r V\right)-   V'   \label{eq:eqhnoA'}\eeq
with vanishing left-hand side.
Light moves radially as $dr_\pm /dt  = \pm A e^{\delta}  $, so time moves faster where $ A e^{\delta}$ is larger.
The criterion of eq.\eq{Thetaspherical} for an apparent horizon becomes $\Theta =-A^2e^{2\delta}$ in static coordinates:
a horizon appears where/when time freezes. 

%The criterion for horizon formation is $\Theta=-\tilde A^2$.

\smallskip

By writing $A(t,r)\equiv 1-2GM(t,r)/r$, the equation for $A$ simplifies into an intuitive
equation for the mass $M$ enclosed in radius $r$:
\beq M' = 4\pi r^2 \left(V + A\frac{h^{\prime 2}}{2}+e^{-2 \delta }  \frac{\dot{h}^2}{2 A}\right).\eeq
If $M(t,0)=M_{\rm in}\neq 0$ a black hole is present, and one needs to solve the equations only outside its horizon.
Convenient boundary conditions for $\delta$ are $\delta(t,0)=0$ or $\delta(t,\infty)=0$:
the two coordinate choices describe the same physics, being related by some redefinition of the time coordinate. 
We follow the standard $\delta(t,\infty)=0$, so that numerical solutions slow down 
before hitting the singularity.
%$t=f(t')$
%Without loss of generality one can assume for $\delta$ a boundary condition 
%at the same point $\delta(t,r=0)=0$.
%(The alternative less convenient choice $\delta(t,r=\infty)=0$ corresponds to ).

For a constant field $h$, the solution $M = M_{\rm in}+4\pi r^3 V/3$ i.e.\ $A = 1 -8\pi G V r^2/3$
reproduces the well known static dS (for $V>0$) and AdS (for $V<0$) solutions,
\beq A = 1 - \frac{2GM_{\rm in}}{r}  - \frac{8\pi G}{3} V r^2 ,\qquad \delta=0,
\eeq
with Hubble rate $\HI^2 = 8\pi G V/3$ if $V>0$. 
For $M_{\rm in}=0$ one has $\Theta=-A^2=0$ at $r=1/\HI$: this is the usual dS horizon.  
In the static limit, the static deSitter coordinates (denoted as $t_{\rm st},r_{\rm st}$ in the equation below)
cover partially the flat FRW coordinates 
(denoted as $t_{\rm FRW}, r_{\rm FRW}$) that, in turn, cover partially the full de Sitter space. Their explicit connection is
\beq t_{\rm st} =t_{\rm FRW}- \frac{1}{2\HI} \ln (1-\HI^2 r_{\rm FRW}^2 e^{2\HI  t_{\rm FRW}}),\qquad  r_{\rm st} = r_{\rm FRW} e^{\HI t_{\rm FRW}}.\eeq
The connection between FRW-like and static-like coordinates is more complicated.

\begin{figure}[t]
$$\includegraphics[width=0.45\textwidth]{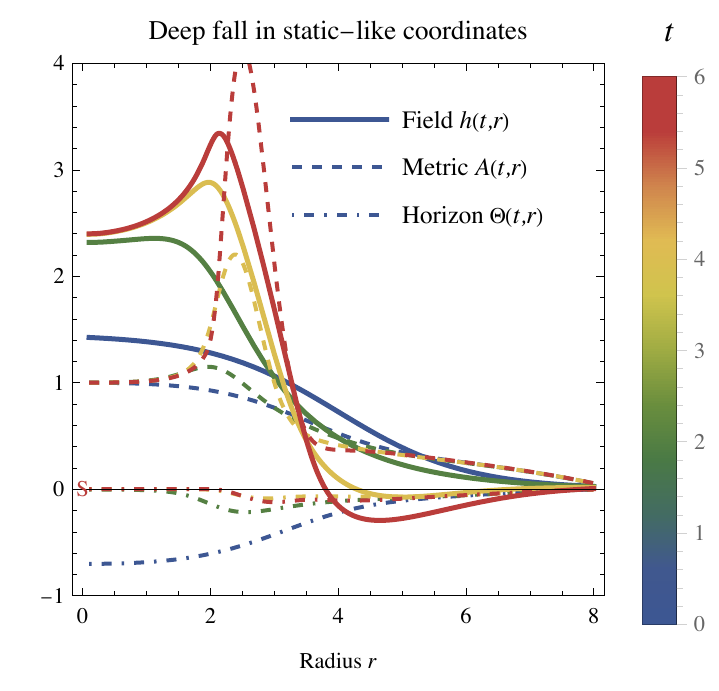}\qquad \includegraphics[width=0.45\textwidth]{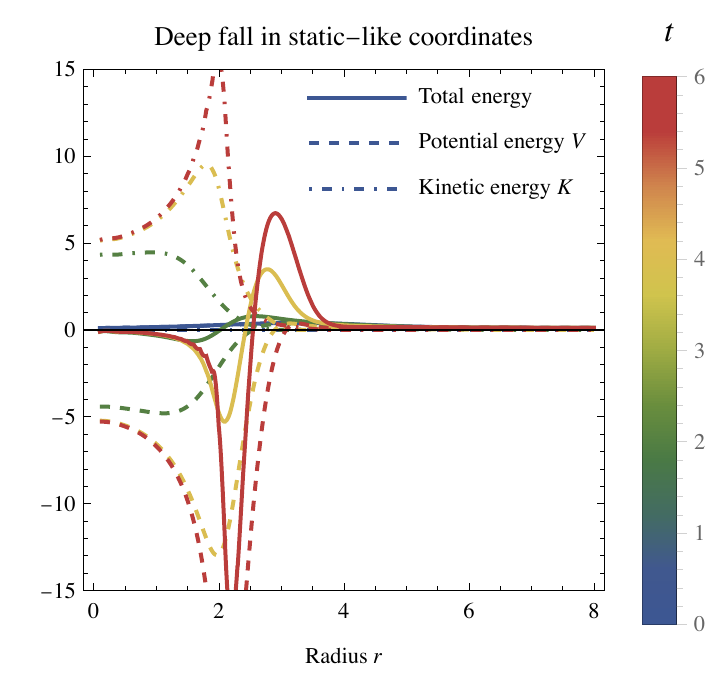}$$
\caption{\it Evolution in Einstein gravity of a scalar bubble beyond the top of the potential.
We work in units in which $h_{\rm top}=1$. 
{\bfseries Left}: the field $h(t,r)$, the metric factor $A(t,r)$ and
the criterion for horizon formation $\Theta(t,r)=-A^2e^{2\delta}$.
%A black-hole horizon appears at the point denoted as {\rm H}. 
{\bfseries Right}: the energy densities. 
\label{fig:DeepFallStatic}}
\end{figure}

\subsubsection*{Numerical simulations in static-like coordinates}
In view of their hybrid nature, we solve numerically eq.s~(\ref{sys:Adelta}) in two independent ways:
with routines built in {\sc Mathematica}, or using the implicit Euler method at 2nd order in the discretisation step.

In static coordinates, even an initial configuration $h_0(r)$ that is $r$-independent acquires a dependence on $r$,
as the field evolves in time.
This can be seen by simply considering a fixed background: the scalar equation becomes $\ddot h = - Ae^{2\delta} V'$, with no Hubble friction in time.
Time evolution of a scalar in a fixed dS background happens faster at smaller $r$; 
while time evolution in a fixed AdS background happens faster at larger $r$.
An initially thin wall does not remain thin because of this effect, in addition to the expansion of the wall.
In flat space ($Ae^{2\delta}=1$), deep fall in a quartic scalar potential happens in a time $\tau \sim 1/h_0\sqrt{-\lambda}$.
%maybe order one factors are $\pi/h_0\sqrt{-2\lambda}$.
This fall time is parametrically the same as the minimal radius $r_0 \sim h_{\rm top}/\sqrt{V_{\rm top}}$ needed for an expanding bubble,
so that its profile gets significantly distorted.
% for expansion,
%so bubbles do not have time to expand during the fall.

%Considering 
%$h(r)$ does not remain constant because . 
%At the origin $A=1$ 
%
%Rather, points where $h>h_{\rm top}$ is above the barrier fall down to the true minimum, while points before the barrier fall 
%to the false minimum.

We assume an initial condition like in eq.\eq{esempio}, except that now $\dot h=0$ with respect to the static-coordinates time
and $\HI=h_{\rm top}/10$.
During the subsequent fall, the kinetic energy density $K$ grows,
while the potential energy $V$ decreases,
and the total energy density $\rho(r) \approx K(r)+V(r)$ tends to vary less,
in view of the absence of Hubble friction.
%In static-like coordinates the total energy density $E(r)$ decreases becoming a bit negative in the central region.
There is no contradiction with the FRW picture of the previous section,
%(which is adapted to the framework of \cite{Riotto}), 
as the energy density
(the time-time component of the energy-momentum tensor $T_{\mu\nu}$) is coordinate-dependent.

The metric factor $\delta(t,r)$ was initially $|\delta|\ll 1$ (under the assumption 
$h_{\rm top}\ll \bP$).
If the potential is shallow the numerical evolution reaches the true minimum $h=h_{\rm min}$; next the scalar $h$ bounces,
leading to an expanding bubble while $\delta$ becomes mildly negative in the interior,
meaning that `time' of the present coordinates runs slower in the interior.

Fig.\fig{DeepFallStatic} shows a solution for a deep scalar fall in an ideally bottom-less potential:
$\delta$ becomes largely negative in the highly-curved inner region,
so that `time' freezes in the interior while the fall is proceeding towards the singularity.
The field falls even more around the border at $r\circa{<} r_0$.
The metric factor $A$ develops around the wall a shape qualitatively similar to fig.\fig{finfoutsample}a, 
indicating that an AdS-like interior develops in the initial dS-like space.
Far away, the metric factor $A(t,r)$ and thereby the enclosed mass $M(t,r)$ roughly keep their initial values,
confirming that far-away quantities don't have time to evolve.
This phenomenon, anticipated in the toy model of section~\ref{Bondi}, is relevant also for the scalar system.
Once again, unlike in the thin-wall idealisation, the numerical solution shows that
the scalar field $h$ extends in the outer region so that an expanding bubble remains despite the inner gravitational collapse.

\section{Black holes with Higgs hair}\label{hair}
Finally, we elaborate on the post-collapse configuration 
that leaves an expanding black hole surrounded by Higgs hair mildly above the instability scale $h_{\rm top}$.
We here show that this configuration cannot stabilise, as static black holes with Higgs hair are unstable.

\smallskip

The scalar equation on a fixed Schwarzschild background, $\delta=0$ and $A= 1- 2GM_{\rm in}/r$ in static coordinates, is
\beq \label{eq:hstaticfixed}h'' + \left(\frac{1}{r}+\frac{1}{r-2 GM_{\rm in}}\right) h' =\frac{V'}{1-2GM_{\rm in}/r} .\eeq
For $V=0$ the scalar equation is solved by $h \propto \ln (1-2GM_{\rm in}/r)$ that diverges 
on the horizon $r_{\rm hor} = 2 G M_{\rm in}$,
together with the energy-momentum tensor.
So, this scalar profile unavoidably modifies the background.
%A positive potential do not qualitatively changes the picture,
%Adding a potential does not seem to qualitatively change the picture
%(an unstable potential $V = - \lambda h^4/4$ adds its own runaway;
%a stable potential makes  the flat-space solution oscillatory, as usual for the modes).
%Since the scalar $T_{\mu\nu}$ also diverges, it modifies the background.
On the other hand, black holes with scalar hair exist 
(even in a fixed Schwarzschild background, see~\cite{1504.08209} for a review)
in the case we are considering: the potential $V(h)$ decreases after a barrier,
and the negative $V'<0$ at $h > h_{\rm top}$ allows
for a cancellation between the two terms in eq.\eq{hstaticfixed} that diverge on the horizon.
A regular solution with $h'_{\rm hor} = r_{\rm hor}  V'(h_{\rm hor} )$ can exist.
In such a case, the fixed Schwarzschild background is a valid approximation for black holes with sub-Planckian hair, 
$h_{\rm top}\ll \bP$.
Since we already have the full equations, we can allow for a generic background.
%Such BH have higher gravitational mass than Schwatzild BH, so they are unstable.
%As $h$ goes between two minima, it's somehow like a bounce.
%Is BH like finding Coleman-deLuccia bounces, as time does not matter?
%The final state could be a BH. This means a boundary condition at $R_{\rm Sch}$
%such that Higgs wave can only enter. 
%Maybe it makes sense to first study what this means for a heavy black hole,
%that can be approximated as static,
%and see if/how the Higgs affects the wall outside.
%A good starting point would be finding the static but unstable solution
%(like the top of Tetradis potential).
%So it's connected to solutions that can go either in or out,
%showing that going out type.
Employing the static-like coordinates of eq.\eq{dsAdelta}, the classical equations~(\ref{sys:Adelta}) 
reduce in the time-independent limit to
%a gravitational background
%\beq ds^2 = A e^{2\delta} dt^2 - \left[\frac{dr^2}{A}+ r^2 (d\theta^2 +d\varphi^2  \sin^2 \theta )\right]\eeq
%where $A$ and $\delta$ can depend on $r,t$.
%\subsection{Scalar static equation}
\beq \label{eq:hstat}
h'' + h' \left(\frac{2}{r}+\frac{A'}{A} +\delta' \right) =  \frac{V'(h)}{A},\qquad
M' = 4\pi r^2 \bigg(V + A\frac{h^{\prime 2}}{2}\bigg),\qquad \delta'= 4\pi G r h^{\prime 2},
\eeq
where $A(r) = 1- 2G M(r)/r$.
These equations agree e.g.~with~\cite{gr-qc/0301062,1606.04018}.
Eliminating $\delta'$ and $M'$, the scalar eq.\eq{eqhnoA'} becomes
\beq h'' +h'   \left(\frac{2}{r}+\frac{2G}{r^2} \frac{M - 4\pi r^3 V}{1-2GM/r}  \right) =  \frac{V'}{1-2GM/r}.\eeq
With these full equations, a black hole hair regular at the horizon exists if the two terms proportional to the divergent $1/A$ term cancel,
implying the boundary condition at the horizon  $2GM(r_{\rm hor})= r_{\rm hor}  $: 
\beq  h'_{\rm hor} = \frac{r_{\rm hor}  V'(h_{\rm hor} )}{1-8\pi G r_{\rm hor} ^2 V(h_{\rm hor} )}.\eeq
This shows that the extra denominator can be neglected in the sub-Planckian limit.
%
%means that the potential must go over a barrier. Numerical solutions find it also goes below $V_{\rm false}$.
Furthermore we are interested in solutions that reach the false vacuum at  $r\to\infty$
\beq h(\infty)=h_{\rm false},\qquad \delta (\infty)=0.\eeq
%where the boundary for $\delta$ is just a convention.
%Since the equation for $\delta$ is decoupled from the other equations, 
The differential equation for $h$ can be solved
starting from an arbitrary $r=r_{\rm hor} =2GM_{\rm in}$
and undershooting/overshooting until reaching $h_{\rm false}$ at $r\to \infty$.
This is similar to computing a vacuum-decay bounce, 
and indeed the solution also describes thermal tunnelling in a black hole background~\cite{1606.04018}.
In the limit of vanishing black hole mass, $r_{\rm hor} =0$, it reduces to the usual bounce for thermal vacuum decay.

\begin{figure}[t]
$$\includegraphics[width=0.45\textwidth]{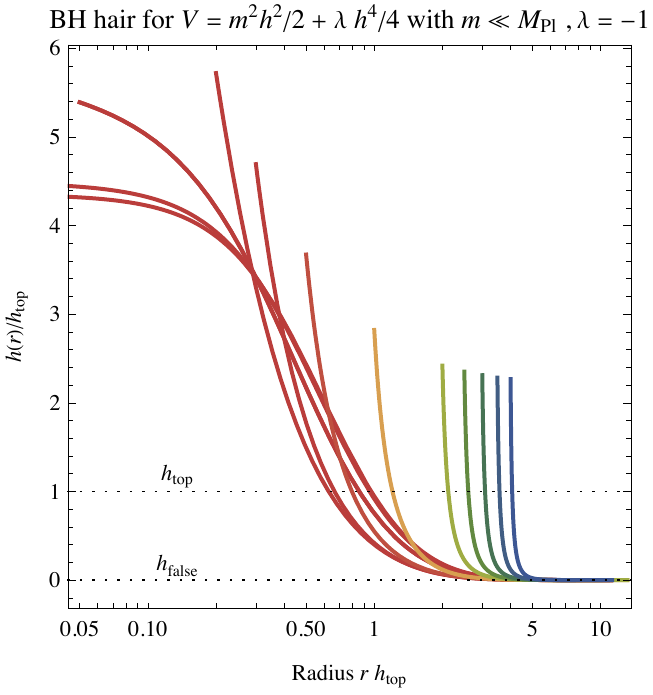}\qquad \includegraphics[width=0.45\textwidth]{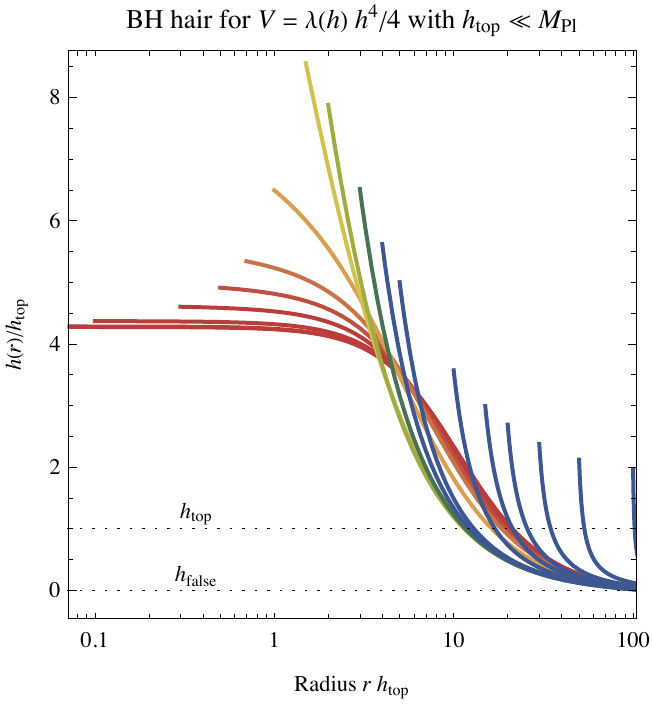}$$
\caption{\it Scalar hair around black holes with different mass $M_{\rm in}=r_{\rm hor}/2G$.
The left plot uses a simple quartic scalar potential;
the right plot uses the SM Higgs potential of eq.\eq{VhSM}.
Solutions only exist for $h_{\rm hor}$ comparable to the instability scale $h_{\rm top}$ of the potential.
Their time evolution shows that in all cases these solutions are unstable.
\label{fig:BHhairsQuarticV}}
\end{figure}

The full eq.s\eq{hstat} show that, in the sub-Planckian limit $h_{\rm top}\ll M_{\rm Pl}$,
the solutions have a length scale $r \sim 1/h_{\rm top}$ so that $|\delta|\ll 1$
and $M'$ are negligible, and the single eq.\eq{hstaticfixed} is sufficient.
Fig.\fig{BHhairsQuarticV} shows numerical solutions in the case of a quartic potential (left)
and the SM Higgs potential (right).
In both cases the solutions for different black hole masses show
a maximal value of the scalar field outside the horizon comparable to $h_{\rm top}$.
The same phenomenon was found in the previous section, when computing the dynamical process
that lead to evolving black holes with scalar hair.

%The scalar vev is maximal $h \circa{<} \hbox{few}\times h_{\rm top}$ for BH with horizon $r \sim 1/h_{\rm top}$

Static black holes with hair are unstable.
This was shown in a specific case 
e.g.~in~\cite{gr-qc/0301062} by adding infinitesimal perturbations and performing the 
stability analysis.
Having the full general-relativistic equations we can see this instability in action, by computing  the time evolution.
The scalar starts getting higher or lower at a radius away from the horizon,
while it evolves slower closer to the horizon (since time is `frozen' there).
In both cases the black hole loses its hair, evolving either into a black hole in the true vacuum
(thereby behaving as an expanding bubble such that $h=h_{\rm min}$ outside)
or into a black hole in the false vacuum 
(thereby behaving as a contracting bubble such that $h=h_{\rm false}$ outside).
As expected, what is found is the typical behaviour of sub-critical or super-critical
bubbles near the critical unstable configuration that describes vacuum decay.
In this language, the solutions found in the past section are super-critical bubbles,
so that their expansion is not surprising.

\myomit{In the limit where the scale of the potential is much sub-Planckian, $m\ll  M_{\rm Pl}$ or equivalently $G\ll 1/m^2$,
the problem simplifies.
\begin{itemize}
\item Since $\delta'=4\pi G r h^{\prime 2}$ one has $\delta=0$.

\item Furthermore the `best' BH have $r_{\rm hor}  \sim 1/m$ i.e.\ large $M$.
Since $ M' = 4\pi r^2 (V + A\sfrac{h^{\prime 2}}{2})$,
$M(r)$ is nearly constant: 
$M'$ is negligible relatively to $M$, but still non-vanishing as needed to satisfy no-hair theorems.
Furthermore at large $M$  one can neglect the $V$ term in the equation for $h$ and the 
only equation simplifies to 
\beq  h'' + h' 
\left[\frac{2}{r}+\frac{2GM}{r(r-2GM)}\right]
=\frac{V'}{1-2GM/r}.\eeq
\end{itemize}
At large $r$ gravity is negligible and the solution is $h\propto e^{-m r}$ where $V \simeq m^2 h^2/2+\cdots$.}

\section{Conclusions}\label{concl}
We studied scenarios that may be realized if the Higgs field or some other scalar 
has a potential with a 
false minimum and a very deep true minimum of negative energy density, 
separated by a potential barrier.
We are interested in the evolution of the system if the field, starting from the
false vacuum, finds its way beyond the potential barrier
$h \circa{>} h_{\rm top}$ 
within some region of space large enough 
so that it can start rolling down the potential towards the true minimum.
Does this process go on until engulfing all space, or can gravity stop it?
The second possibility was supported by a thin-wall calculation in~\cite{Riotto}.
However, we found that the thin-wall approximation is not applicable 
before the field reaches its minima.

In order to settle the issue, we performed a full computation in the
context of Einstein gravity with a  scalar.
We found that the negative potential energy starts an accelerating 
gravitational collapse
that results in a central singularity, but does not stop the expanding scalar bubble.
%This can also be seen as the balding process of a black hole that 
%temporarily developed a scalar hair
%during the transition from the false to the true vacuum.
The bubble expands at the speed of light, while its potential energy 
has the spatial profile typical of a  `sinkhole',
with a central singularity located behind an apparent horizon.
The configuration is time-dependent, as the central region 
continuously accretes energy.

In summary, this is the naively expected result: 
regions with  $h \circa{>} h_{\rm top}$ and size $r_0 \circa{>} 1/h_{\rm top}$ can fall 
towards the deep true minimum, and find a way to fall.

% the deep state does allow to say that  sinkholes are dubbed `portal to hell' 

\smallskip

In particular, this scenario applies to the case of the Higgs boson, 
the only scalar discovered so far.
Current best-fit values indicate that
the potential of the Standard Model Higgs field becomes negative when extrapolated to 
ultra-large field values.
The renormalized Higgs quartic coupling in 
$V \approx \lambda(h) h^4/4$ turns negative around 
$h_{\rm top} \sim 10^{10-11}\GeV$,
even though, taking $\pm 3\sigma$ uncertainties into account, 
the instability can be pushed above the Planck scale or disappear~\cite{1307.3536}.
If more accurate future measurements confirm this instability,
the minimal energy of a Higgs field configuration that can trigger 
a catastrophic process that would destroy the universe is estimated as
\beq E \sim r_0^3 V(h_{\rm top}) \circa{>} h_{\rm top} \sim 1\,{\rm Joule}.\eeq
Packing energy into the Higgs field in a sufficiently
small region is far beyond current technological limits. 
The scenario is relevant for the evolution of the early universe, and implies 
bounds on the scale of processes that can create sufficiently strong fluctuations of
the Higgs field, or any other field with a potential with a very deep minimum~\cite{
Higgstory,1112.3022,1210.6987,1301.2846,1503.05193,1605.04974,1606.00849,1608.02555,1608.08848,1706.00792,1809.06923,2011.03763}.

%A particle scattering with $\sqrt{\hat{s}}\sim h_{\rm top}$ or larger
%would trigger such instability with exponentially suppressed probability, according to the %computations in~\cite{Rubakov:1991fb,hep-ph/9703256,hep-ph/9704431,hep-ph/9910333}.
%This has not been tested, as the collision with highest energy known to have ever occurred in our %past light-cone is 
%estimated as $\sim 10^{11-12}\GeV$, among ultra-high-energy cosmic rays~\cite{Hut:1983xa}. 
%With current technology, a collider needs to be planet-size to reach the possible instability %scale.
%Current colliders produce Higgs bosons with lower energy $E \sim M_h$, that would only allow to 
%build hypothetical bubbles larger than $r_0 \sim 1/M_h$. In order to trigger the instability, such %`large' bubbles 
%would need $N \sim (h_{\rm top}/M_h)^3\sim 10^{24}$ packed Higgs bosons, before they decay.
%%$E \sim h_{\rm top}^4/M_h^3 \sim 10^{24}\,{\rm J}$ (not much above the yearly energy production)
%This is technologically unrealistic. 

\small\frenchspacing
\paragraph{Acknowledgments}
We thank V. De Luca, A. Kehagias, A. Riotto for discussions on the subject
and J. Rizos,  D. Teresi for advice on the numerics.

\end{document}